\documentclass[conference]{IEEEtran} 		
\IEEEoverridecommandlockouts
\usepackage{ushort}
\usepackage[mathcal]{euscript}
\usepackage{amsmath}
\usepackage{amsthm}
\usepackage{amssymb}
\usepackage{graphicx}
\usepackage{epsfig}
\usepackage{booktabs}
\usepackage{colortbl}
\usepackage{amsmath}
\usepackage{xcolor}
\usepackage{graphicx}
\usepackage{algpseudocode}
\usepackage{algorithmicx}
\usepackage{algorithm}
\usepackage{cite}
\usepackage{rotating}
\usepackage{array}
\def \aed {\lambda_{\text{ed}}}
\def \apu {\lambda_{\text{pu}}}
\def \bed {\beta_{\text{ed}}}
\def \bpu {\beta_{\text{pu}}}
\def \med {\mu_{\text{ed}}}
\def \mpu {\mu_{\text{pu}}}

\def \sed {s_{\text{ed}}}
\def \spu {s_{\text{pu}}}
\def \ued {U_{\text{ed}}}
\def \upu {U_{\text{pu}}}

\begin{document}

\title{An \emph{Online} Delay Efficient Packet Scheduler for M2M Traffic in Industrial Automation}

\author{\IEEEauthorblockN{Akshay Kumar, Ahmed Abdelhadi and Charles Clancy}
\IEEEauthorblockA{Hume Center, Virginia Tech\\
Email:\{akshay2, aabdelhadi, tcc\}@vt.edu}
\thanks{This research is based upon work supported by the National Science Foundation under Grant No. 1134843. 
}
}

\maketitle

\begin{abstract}
Some Machine-to-Machine (M2M) communication links particularly those in a industrial automation plant have stringent latency requirements. In this paper, we study the delay-performance for the M2M uplink from the sensors to a Programmable Logic Controller (PLC) in a industrial automation scenario. The uplink traffic can be broadly classified as either Periodic Update (PU) and Event Driven (ED). The PU arrivals from different sensors are periodic, synchronized by the PLC and need to be processed by a prespecified \emph{firm} latency deadline. On the other hand, the ED arrivals are random, have low-arrival rate, but may need to be processed quickly depending upon the criticality of the application. To accommodate these contrasting Quality-of-Service (QoS) requirements, we model the utility of PU and ED packets using step function and sigmoidal functions of latency respectively. Our goal is to maximize the overall system utility while being proportionally fair to both PU and ED data. To this end, we propose a novel \emph{online} QoS-aware packet scheduler that gives priority to ED data as long as that results the latency deadline is met for PU data. However as the size of networks increases, we drop the PU packets that fail to meet latency deadline which reduces congestion and improves overall system utility. Using extensive simulations, we compare the performance of our scheme with various scheduling policies such as First-Come-First-Serve (FCFS), Earliest-Due-Date (EDD) and (preemptive) priority. We show that our scheme outperforms the existing schemes for various simulation scenarios.
\end{abstract}


\section{Introduction}
Automation for industrial monitoring and control processes has become commonplace these days. It can be modeled as a star-topology Machine-to-Machine (M2M) network of various process sensors communicating to a Programmable Logic Controller (PLC) on the \lq uplink channel\rq~ which then feeds back the output to a control device (\lq downlink channel\rq) to ensure the desired process operation. Due to increasing complexity and scale of modern industrial monitoring and control systems, there has been a paradigm shift from the use of point-to-point wired systems to wireless technologies for both uplink and downlink communication. Wireless technologies specifically tailored for signaling in industrial monitoring systems, such as WISA, Zigbee Pro etc. (which are based on IEEE 802.15.1 and IEEE 802.15.4 standard) are becoming increasing popular (see \cite{Zand12} and references therein).

Typically, the industrial M2M traffic has very low latency requirements (of order of few milliseconds). It can be broadly categorized as either \emph{non real-time} (no deadline for task completion) or \emph{soft real-time} (decreased utility if deadline not met) or \emph{firm} (zero utility if deadline not met) or \emph{hard} (system failure if deadline not met). For instance, the messages from a instrumented protective system or safeguarding system have \emph{hard} real-time Quality-of-Service (QoS) requirements whereas preventive maintenance applications are typically served in \emph{non real-time}. With increase in network size, the available computational and (wireless) communication resources gets shared among a large number of sensors. This makes it harder to provide real-time QoS due to simultaneous access attempts from multiple sensors \cite{3GPPreport} and also wastage of wireless spectrum continuously allocated to sensors with very low transmission duty cycle \cite{Drajic12}.

Therefore in this paper, we develop an \emph{online} delay-efficient packet scheduler for the M2M traffic. To begin, we first use source M2M traffic model in \cite{Navid13} to classify the sensor data into Periodic Update (PU) and Event Driven (ED) packets. The PU arrivals are periodic with \emph{firm}\footnote{We assume \emph{firm} QoS for PU instead of \emph{hard} QoS because typically the periodic data is a slowly varying function of time and will not change drastically from one period to another. So if we don't meet the completion deadline for a PU task, the PLC can use an estimate for the PU based on the recent history, albeit with some degradation in utility. The estimate will typically be better if we store more historical PU data but this increase the memory size and computational complexity.} latency deadline while the ED arrivals are random without hard service deadlines. 

\emph{Paper Contributions}: We map the contrasting QoS requirements for PU and ED packets using step and sigmoidal utility functions. We define the overall system utility as the product of (time) averaged PU and ED utilities so as to ensure proportional fairness between the two classes. Our goal is to determine the optimal scheduling policy that maximizes the proposed system utility. However, due to the randomness in arrival and service times, there exists no delay-optimal \emph{online} algorithm \cite{TiaLiuShankar}.

Therefore, we propose an \emph{online} delay-efficient heuristic scheduler for the uplink traffic at the PLC, that  gives priority to ED data as long as that does not result in a utility loss for the PU data. This results in a higher utility for the ED data. However, at high server utilization, as the PU task start failing their task completion deadlines, we propose to drop such PU packets as their completion does not result in utility. This helps us to mitigate overall congestion and thus allows us to meet the deadlines of remaining PU tasks and also serve the ED data with higher utility. Using extensive simulations, we show that this results in an overall increase in system utility as compared to other popular scheduling policies such as First-Come-First-Serve (FCFS), Earliest Due Date (EDD), priority scheduling etc. Lastly, the proposed scheduler is agnostic to the wireless technology being used for the M2M uplink and is also independent of the hardware-software architecture used in practical real-time embedded systems. Also it can easily adapt to accommodate time-varying arrival and service rates for the PU and ED packets. 

\emph{Paper Outline}: The rest of the paper is organized as follows. Section~\ref{related} details the related work. Section~\ref{sysModel} introduces the system model and defines the utility functions for PU and ED data. Then in Section~\ref{probForm}, we formulate the utility maximization problem and in Section \ref{propSch} describe the proposed scheduling scheme at PLC. Section~\ref{Results} presents simulation results. Finally Section~\ref{concl} draws some conclusions.

\section{Related Work}
\label{related}
Over the last decade, numerous hardware and software technologies including embedded real-time systems, Ethernet, Low-power Wi-Fi, Zigbee have made their way into industrial automation platforms so as to increase their flexibility and efficiency. However, this has resulted in increased complexity both at the network and component level, due to sharing of physical resources for various applications. This makes it harder to guarantee QoS for (hard) real-time applications. It was shown by Buttazzo in \cite{Buttazzo04} that merely increasing the computational resources is not too useful unless appropriate scheduling strategies are put in place. Consequently, a number of prior works have looked into integrating the design of real-time scheduling schemes in middle-ware (such as CORBA) based architectures. Tommaso et. al. in \cite{Tommaso09} presented a real-time Service Oriented Architecture (SOA) for industrial automation. However, all of these works are mostly tangential to our line of work in the sense that our proposed scheduler is agnostic of the hardware-middleware-software architecture and maps the real-time QoS requirements of M2M traffic into utility functions that we aim to maximize.    

Another line of work focuses on QoS-aware packet scheduler for M2M traffic in Long Term Evolution (LTE) network (see \cite{Gotsis12} and references therein).  Most of these works use some variants of Access Grant Time Interval scheme for allocating fixed or dynamic access grants over periodic time intervals to M2M devices. Nusrat et. al. in \cite{Afrin13} designed a packet scheduler for M2M in LTE so as to maximize the percentage of uplink packets that satisfy their individual budget limits.  Ray and Kwang in \cite{Ray13} proposed a distributed congestion control algorithm which allocates rates to M2M flows in proportion of their demands. 
Unlike our work, all of these works design packet scheduler specific to a wireless standard such as LTE and are thus heavily influenced by the Medium Access Control (MAC) architecture of LTE. Unlike our work, they also don't explicitly segregate data arrivals into different QoS classes.

Lastly, a number of scheduling algorithms have been proposed specifically for real-time embedded systems (see \cite{Buttazzo11} and references therein) that are agnostic to the application scenario, wireless technology used or hardware-software architecture. These schemes assume hybrid task sets comprising of \emph{hard} periodic requests and \emph{soft real-time} aperiodic requests. They are broadly classified into Fixed-priority and Dynamic priority assignments. The goal of all these schemes is to guarantee completion of service (before deadline) for all periodic request and simultaneously aim to reduce the average response times of aperiodic requests.  Fixed priority schemes schedule periodic tasks using Rate-Monotonic algorithm but differ in service for aperiodic tasks. Dynamic scheduling algorithms schedule periodic tasks using EDD scheme and allow better processor utilization and enhance aperiodic responsiveness as compared to the fixed priority schemes. The drawbacks of these schemes is that they assume that the random service times for each tasks are known \emph{apriori}. Hence these schemes are not truly \emph{online} in their present form and would require considerable modifications.

\section{System Model}
\label{sysModel}
\begin{figure}%
\centering
\includegraphics[width=\columnwidth]{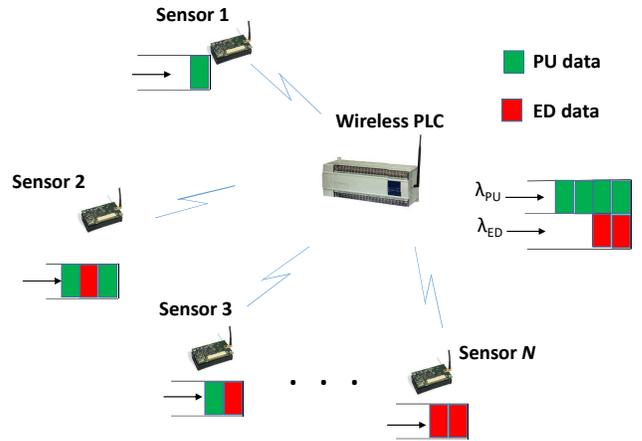}%
\vspace{0pt}
\caption{System Model.}%
\label{systemModel}%
\end{figure}

Fig~\ref{systemModel} shows the system model illustrating the queuing process in the uplink channel (i.e., from sensors to the PLC) in an industrial automation setting. Each of the $N$ sensors transmits two types of packets to the PLC: frequent periodic updates (PU) for some process related measurement or sporadic event-driven (ED) packets triggered when the sensor measurement for a physical quantity crosses its threshold. The PU arrivals from the $i^{\text{th}}$ sensor is modeled as a deterministic periodic process with period $T_{\text{pu}}^i$ while the ED arrivals is modeled as a Poisson process with rate $\aed^i$. Therefore the queuing process at the PLC has net arrival rate $\apu$ and $\aed$ for PU and ED packets respectively, where
\begin{align}
\apu = \sum_{i=1}^{N} 1/T_{\text{pu}}^i~\text{and}~ \aed = \sum_{i=1}^{N} \aed^i.
\end{align}

Let $\spu$ and $\sed$ denote the size for PU and ED packets. The service time for PU and ED packets is assumed to be exponential with rate $\mpu$ and $\med$ respectively given by
\begin{equation}
\mpu = \mu/\spu~\text{and}~\med = \mu/\sed,
\label{serviceRate}
\end{equation} 
where $\mu$ is the service rate at PLC per unit packet size. In general, the total time spent by a packet in the system, $T$, can be written as sum of following components,
\begin{equation}
T = T_{\text{trans}}+T_{\text{prop}}+T_{\text{cong}}+T_{\text{queue}}+T_{\text{ser}},
\label{sojTime}
\end{equation}
which denote the following component delays:
\begin{itemize}
	\item $T_{\text{trans}}$: Transmission delay at the sensor.
	\item $T_{\text{prop}}$: Propagation delay from the sensor to PLC.
	\item $T_{\text{cong}}$: Congestion delay due to shared wireless channel in large-scale sensor network.
	\item $T_{\text{queue}}$: Queuing delay at the PLC.
	\item $T_{\text{ser}}$: Processing time for a packet at the PLC.
\end{itemize}
In this work, we ignore all the terms except the queuing delay at PLC $T_{\text{queue}}$ and the service time $T_{\text{ser}}$. This is because the packet size is usually small enough to ignore $T_{\text{trans}}$ and the sensors and PLC are usually close enough on the shop floor to permit the usage of low power wireless signaling and thus ignore $T_{\text{prop}}$. Also, we assume that the available wireless spectrum is large enough to allocate a dedicated transmission channel to each sensor; so there is no congestion delay $T_{\text{cong}}$.

At the PLC, the PU and ED packets form two separate queues due to different attributes such as packet size and latency requirements. However, for simplicity for a given PU or ED class, we don't distinguish between packets arriving from different sensors. Now the queuing delay for each class depends upon the scheduling policy adopted at the PLC. The scheduling policy at PLC should be chosen as to maximally satisfy the latency constraints of PU and ED packets. 

\section{Problem Formulation} 
\label{probForm}
\begin{figure}
\centering
   \includegraphics[width=2.3 in,height=1.7 in]{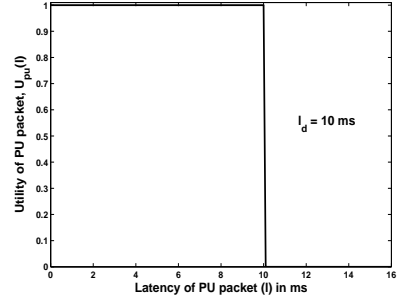}
   \label{puUtility} 
\caption{\footnotesize{Utility function for PU packet.}}
\end{figure}

\begin{figure}%
\centering
\includegraphics[width=2.3 in,height=1.7 in]{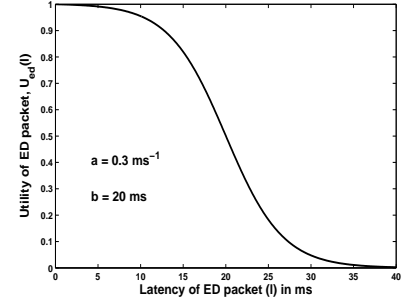}%
   \label{edUtility} 
\caption{\footnotesize{Utility function for ED packet.}}
\end{figure}

We first map the latency requirements onto the utility functions for both PU and ED classes as shown in Fig.~\ref{puUtility} and~\ref{edUtility}.
\subsection{Utility Functions}

\subsubsection{PU utility} As stated previously, the PU packets need to be processed within a prespecified (usually by PLC) time interval at the end of which there is no utility in serving the packet. So, we define PU utility as,
\begin{equation}
\upu(l) = \begin{cases} 1 & \text{if } l < l_d \\
 0 & \text{if } l \geq l_d, \end{cases}
\label{puUtil}
\end{equation}
where $l_d$ is the latency deadline at which utility drops to 0.

\subsubsection{ED utility} Unlike, the PU packets, ED packets do not have a hard deadline, rather their utility is a strictly decreasing function of latency. However the utility does not decrease much until some nominal delay which depends on criticality of application. Also at large latency values, the utility becomes close to 0 and does not change appreciably with further increase in latency. Therefore, we define the ED utility as a sigmoidal function, as in \cite{GhorbanzadehMILCOM2014, AbdelhadiICNC2015}, given by,
\begin{equation}
\ued(l) = 1-c\left(\frac{1}{1+{\text{e}}^{-a(l-b)}}-d\right),
\label{edUtil}
\end{equation}
where, $c=\frac{1+{\text{e}}^{ab}}{{\text{e}^{ab}}}$ and $d = \frac{1}{1+{\text{e}^{ab}}}$. We note that $\ued(0)=1$ and $\ued(\infty) = 0$. The parameter $a$ is the utility roll-off factor whose value depends on the criticality of the application. The inflection point of \eqref{edUtil} occurs at $l = b$ and depends on the average service time (at PLC) for ED packet.

\subsection{System utility function} For a given scheduling policy $\mathcal{P}$, we use utility proportional fairness in our system utility (similar to the system utility introduced in \cite{AbdelhadiMobicom2013, Ghorbanzadeh_arxiv_journal1}) which is given by,
\begin{equation}
V(\mathcal{P}) = {\bold{\upu}}^{\beta_{\text{pu}}}(\mathcal{P})*{\bold{\ued}}^{\beta_{\text{ed}}}(\mathcal{P}),
\label{sysUtil}
\end{equation}
where $\bold{\upu}(\mathcal{P})$ and $\bold{\ued}(\mathcal{P})$ are the utility of PU and ED packets in the steady state given as,
\begin{align}
\bold{\upu}(\mathcal{P}) &= \lim_{{T_{\text{s}}} \rightarrow \infty} \sum_{i=1}^{M_{\text{pu}}(T_{\text{s}})}{\frac{\upu(l_i(\mathcal{P}))}{M_{\text{pu}}(T_{\text{s}})}}, \\
\bold{\ued}(\mathcal{P}) &= \lim_{{T_{\text{s}}} \rightarrow \infty} \sum_{i=1}^{M_{\text{ed}}(T_{\text{s}})}{\frac{\ued(l_i(\mathcal{P}))}{M_{\text{ed}}(T_{\text{s}})}},
\end{align}
where $M_{\text{pu}}(T_{\text{s}})$ and $M_{\text{ed}}(T_{\text{s}})$ are the number of PU and ED packets served in time $T_{\text{s}}$ and $l_i$ is the latency of the $i^{\text{th}}$ packet. The parameters $\beta_{\text{pu}}$ and $\beta_{\text{ed}}$ denote the relative importance of average PU and ED utility. A higher value of $\beta$ implies higher importance of the utility of that class towards the overall system utility.

\subsection{Optimal scheduler} We now describe the utility maximization problem that needs to be solved to determine the optimal scheduling policy at PLC. It is given by,
\begin{align}
\operatorname*{max}_{\mathcal{P}}~V(\mathcal{P}) = {\bold{\upu}}^{\beta_{\text{pu}}}(\mathcal{P})*{\bold{\ued}}^{\beta_{\text{ed}}}(\mathcal{P}). 
\label{optimProb}
\end{align}
If the service times of PU and ED were deterministic or known \emph{apriori}, then the optimal scheme is to schedule PU jobs so that they are completed exactly at the deadline $l_d$. However, since the arrival time of ED packet, the service times for PU and ED packets are random, it is not possible to determine an \emph{online} optimal scheduler \cite{TiaLiuShankar}. Therefore, we propose an \emph{online} heuristic scheduler that aims to maximize the utility function and is described in the next section.

\section{Proposed Scheduler}
\label{propSch}
The utility of PU remains at 1 until the deadline at which it becomes 0. Therefore, it would be best (from the perspective of ED latency) to delay the service of PU as much as possible but ensuring that we serve it before deadline. However, the randomness in PU and ED service times makes it hard to determine the optimal start of service for each PU packet in real-time. To overcome this problem, our proposed scheduler gives priority to ED packets as long as the current latency of PU packet is less than predetermined threshold $l_{t}, (0<l_{t}<l_d)$. If the latency of PU packet exceeds $l_{t}$ and PLC is processing an ED packet, then it gets preempted and the PU packet gets service. This ensures that we reduce latency of ED packets while ensuring that most of the PU packets meet their latency deadline. As we will show later, there exists an optimal value of $l_t$ that maximizes the system utility for given set of system parameters. Using simulations, the optimal $l_t$ can be stored in a Look-Up-Table (LUT) prior to the deployment of proposed scheduler. Thus the proposed scheme can easily \emph{adapt} to changing packet arrival rate, packet size, number of sensor nodes etc. by looking up and using the optimal $l_t$ for each case.

Another novel feature of our scheduler is that we drop PU packets from service or queue that exceed their latency deadline as there is no resultant utility from servicing such packets. However, dropping them from service will reduce congestion and thus reduce latency for PU and ED packets in queue. The proposed scheduler is described in detail in Algorithm~\ref{propScheduler}.
 
\vspace{-5pt}
\section{Simulation Results}
\label{Results}
In this section, we use Monte-Carlo simulations to evaluate the system utility performance of our scheduler against various standard scheduling policies such as FCFS, EDD, Preemptive priority scheduling. The simulation time $T_s$ is set to a large value ($40$~s) to ensure a steady-state queuing behavior. $N$ is set to 50 sensors. For simplicity, we set the PU arrival period from each of the $N$ sensors, $T_{\text{pu}}^i = 50$~ms. The ED arrival rate, $\aed^i$ is same for all sensors and set to $0.0068$/ms. PLC service rate is $\mu=100$~bits/ms. The relative importance of PU and ED is set equal, i.e. $\bpu=\bed=1$, unless mentioned otherwise. The size of PU and ED packets is set to $\spu = 100$~bits and $\sed = 200$~bits respectively and does not change over time or different sensors. The latency deadline for PU, $l_d$ is set to $10$~ms for all sensors. For ED utility function, we set $a=1$/ms and $b=20$~ms unless mentioned otherwise.

\subsection{Impact of latency threshold and ED utility parameter, $b$}
\begin{figure}%
\centering
\includegraphics[height = 2.1 in, width=3 in]{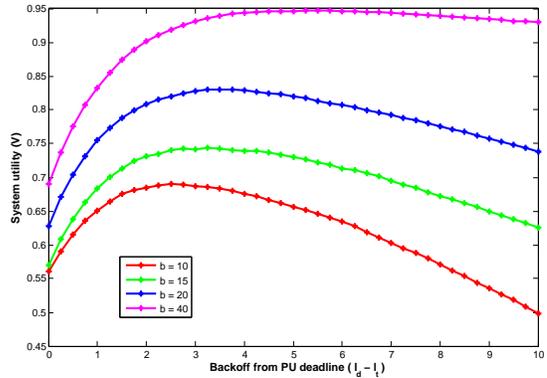}%
\caption{Selection of optimal PU latency threshold $l_t$.}%
\label{varyLatThreshold}%
\end{figure}

Fig.~\ref{varyLatThreshold} illustrates the impact of varying the PU latency threshold $l_t$ in proposed scheduler (without PU drop) for different values of parameter $b$. We note that for a given $b$, there exists an optimal value of $l_t$ that maximizes the system utility. However, at higher values of $b$, the system utility becomes less sensitive to $l_t$ value around the maxima. This is because the ED utility does not vary appreciably with latency when $b$ is high; and PU utility does not vary much at low values of $l_t$.

\subsection{Impact of ED utility parameter, $a$}
\begin{figure}%
\centering
\includegraphics[height = 2.1 in, width=3 in]{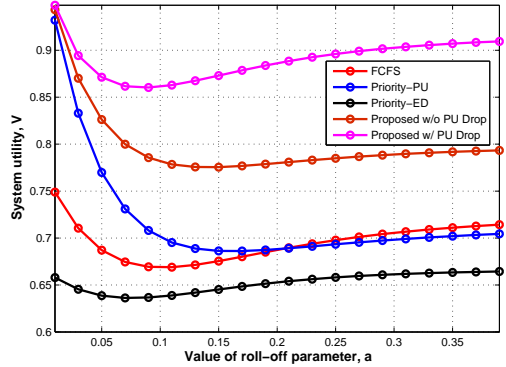}%
\caption{Impact of criticality of ED application set by parameter $a$.}%
\label{varyParamA}%
\end{figure}

Fig.~\ref{varyParamA} illustrates the impact of varying the ED utility parameter $a$ on the system utility for different scheduling policies. As $a$ is increased, ED utility function more brick-walled with the step at latency of $b$. We observe that at very low value of $a$, system utility is very high for the proposed schemes. This is because ED utility remains constant with increase in latency. As $a$ is increased, system utility decreases, attains a minima, increases and then settles down. This is because at high $a$, the ED utility becomes a brick-wall function and any further increase in $a$ does not affect system utility.

\subsection{Impact of number of sensor nodes}
\begin{figure}%
\centering
\includegraphics[height = 2.1 in, width=3 in]{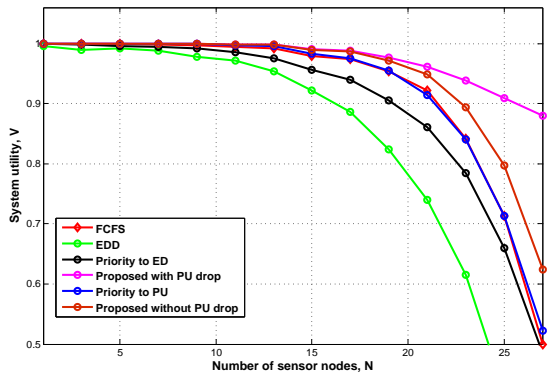}%
\caption{System utility as network size is increased.}%
\label{varyNumNodes}%
\end{figure}

Fig.~\ref{varyNumNodes} shows the impact of increasing number of sensors on the system utility for different scheduling policies. For proposed scheduler, we run Algorithm~\ref{propScheduler} over $0<l_t<l_d$ and determine the optimal $l_t$ that results in maximum utility $V$. Clearly, the proposed scheduler (irrespective of dropping PU packets) always results in a higher system utility compared to all other schemes. However the performance improvement is starkly better as $N$ is increased. This is because at large $N$, the latencies of PU and ED becomes more interdependent and order of service for queued jobs becomes quite important.
Also at large $N$, the proposed scheme with PU drop results in higher dropped packets and thus relieves congestion and results in a significantly higher utility as compared to other schemes.


\subsection{Impact of ED arrival rate}
\begin{figure}%
\centering
\includegraphics[height = 2.1 in, width=3 in]{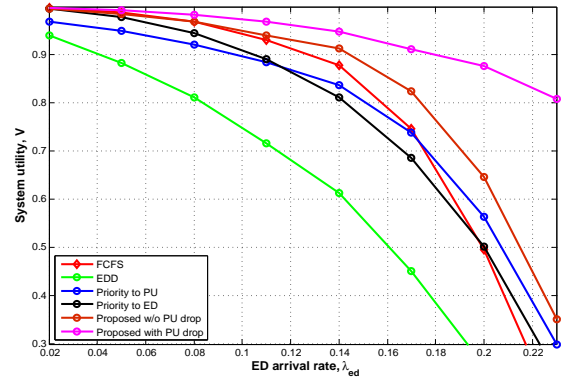}%
\caption{System utility as ED arrival rate is increased.}%
\label{varyEDrate}%
\end{figure}

Fig.~\ref{varyEDrate} shows the impact of increasing the frequency of ED arrivals for a given network. Again for similar reason as stated earlier, we note that the proposed scheduler results in higher utility compared to other schemes. 

\section{Conclusions}
\label{concl}
In this paper, we presented a delay-efficient packet scheduler for uplink M2M traffic. To this end, we classified the uplink traffic as PU or ED packets and mapped their latency requirements onto utility functions. We then proposed a novel packet scheduler to maximize the system utility. We defined the system utility as the weighted product of average PU and ED utility so as to ensure proportional fairness. Using extensive simulations, we did a comparative analysis of the proposed scheme with standard scheduling policies such as FCFS, EDD, Preemptive priority etc. We note that as the network size increases or ED arrival rate increases, there is a stark difference between the utility of proposed scheme and the rest.

\begin{algorithm*}[h]
\caption{Proposed \emph{Online} Packet Scheduler}
\label{propScheduler}
 \textbf{Inputs} \\
  \text{$l_{d}$ and $l_{t}$ are latency deadline and threshold for PU utility such that $0<l_t<l_d$}\\
	\text{$\upu(.)$ and $\ued(.)$ are utility functions for PU and ED packets}\\
	\text{$\bpu$ and $\bed$ are parameters for system utility in \eqref{sysUtil}}\\
	\text{$T_{\text{pu}}^a$ and $T_{\text{ed}}^a$ are arrays for arrival times of PU and ED packets}\\
	\text{$T_{\text{pu}}^s$ and $T_{\text{ed}}^s$ are arrays for service times of PU and ED packets}\\
	\textbf{Outputs} \\
	\text{$V$ is optimal policy}\\
	\textbf{Local} \\
	\text{$T_1$, $T_2$, $T_3$ store current time, service completion time and next arrival time}\\
	\text{$x$, $z$ is type of next packet to be served and the current packet in server. Set them to $1$ for PU and $2$ for ED packet}\\
	\text{$p_1$, $p_2$ are pointers to the next PU and ED packet to be served}\\
	\text{$T_{\text{pu}}^d$ and $T_{\text{ed}}^d$ are arrays for completion times of PU and ED packets}\\
	\text{$D_{\text{pu}}$ and $D_{\text{ed}}$ are arrays for latency of PU and ED packets}\\
	\text{$\bold{U}_{\text{pu}}$ and $\bold{U}_{\text{ed}}$ are average utilities for PU and ED packets} \\	
	\text{$r_1$, $r_2$ are sizes of arrays $T_{\text{pu}}^a$ and $T_{\text{ed}}^a$. Set $\epsilon$ to infinitesimally small value.}\\
	\textbf{Initialization} \\
	\text{$p_1 = 1$, $p_2 = 1$, $T_1 = 0$, $T_2 = 0$}\\
	\textbf{Begin Algorithm:}
		\begin{algorithmic} 
		\label{algoProp}
		\While{$p_1 \leq r_1$ or $p_2 \leq r_2$}
		\If{$T_{\text{pu}}^a(p_1) <= T_{\text{ed}}^a(p_2)$} \Comment{Determine whether next arrival is PU or ED}
		 \State $T_3 = T_{\text{pu}}^a(p_1)$, $x=1$ \Comment{Set next arrival time and packet type to PU}
		\Else
		 \State $T_3 = T_{\text{ed}}^a(p_2)$, $x=2$ \Comment{Set next arrival time and packet type to ED}
		\EndIf
		\If{$T_1 \leq T_3$} 
			\State $T_1 = T_3$ \Comment{Advance current time to next PU or ED arrival} 
		\ElsIf{$z=2$ and $T_2 > T_{\text{pu}}^a(p_1)+l_t$} \Comment{Preempt active ED job if PU latency exceeds $l_t$ }
			\State $p_2 = p_2-1$, $z=1$
			\State $T_{\text{ed}}^s(p_2) = T_2-(T_{\text{pu}}^a(p_1)+l_t)$ \Comment{Residual service time for preempted ED job}
			\State $T_1 = T_{\text{pu}}^a(p_1)+l_t$ \Comment{Retreat current timer to preemption instant}
		\ElsIf{$T_1 > \text{max}(T_{\text{pu}}^a(p_1), T_{\text{ed}}^a(p_2))$} \Comment{Check if both PU and ED jobs present in queue}
			\If{$T_1 > T_{\text{pu}}^a(p_1) + l_t$} \Comment{Service PU job first if its latency exceeds $l_t$}
				\State {$x=1$}
			\Else
				\State {$x=2$}
			\EndIf
		\EndIf
		\If{$x=1$} \Comment{Update the timers and pointers for next PU or ED packet}
			\State {$z=1$, $T_2 = T_1+T_{\text{pu}}^s(p_1)$, $T_{\text{pu}}^d(p_1) = T_2$, $p_1 = p_1+1$}
		\Else 
			\State {$z=2$, $T_2 = T_1+T_{\text{ed}}^s(p_2)$, $T_{\text{ed}}^d(p_2) = T_2$, $p_2 = p_2+1$}
		\EndIf
		\If{$z=1$ and $T_2>T_{\text{pu}}^s(p_1-1)+l_d$} \Comment{Drop the PU packet with latency exceeding the deadline}
			\State {$T_{\text{pu}}^d(p_1-1) = T_{\text{pu}}^a(p_1-1)+l_d+\epsilon$, $T_1 = \text{max}(T_1,  T_{\text{pu}}^a(p_1-1)+l_d)$}
		\Else
			\State {$T_1 = T_2$} \Comment{Advance the timer to service completion time}
		\EndIf
		\EndWhile
		\State {$D_{\text{pu}} = T_{\text{pu}}^d-T_{\text{pu}}^a$, $D_{\text{ed}} = T_{\text{ed}}^d-T_{\text{ed}}^a$} \Comment{Calculate latency for PU and ED packets}
		\State {$\bold{U}_{\text{pu}} = \text{mean}\left(\upu (L_{\text{pu}})\right)$, $\bold{U}_{\text{ed}} = \text{mean}\left(\ued (L_{\text{ed}})\right)$, $V = \bold{U}_{\text{pu}}^{\bpu}*\bold{U}_{\text{ed}}^{\bed}$} \Comment{Determine PU, ED and system utility}
	\end{algorithmic}	
\end{algorithm*}

\bibliographystyle{ieeetr}	

\bibliography{sysCon}	

\end{document}